\documentclass{iopart}
\usepackage{iopams,setstack}

\newtheorem{prop}{Proposition}
\newenvironment{proof}{\textbf{Proof.}}{\hfill$\square$}

\newcommand{\uu}{{\mathbf u}}
\newcommand{\MM}{\mathbf M}
\newcommand{\Eu}{\mathbf E_\uu}
\newcommand{\xx}{\mathbf x}
\newcommand{\II}{\mathbf I}
\newcommand{\RR}{\mathbb R}
\newcommand{\LL}{\mathbf L}

\begin{document}
\title {Poincar\'e covariance of relativistic quantum position}
\author{Sz~Farkas\dag, Z~Kurucz\dag\ddag\ and M~Weiner\dag}
\address{\dag\ Department of Applied Analysis, 
  E\"otv\"os Lor\'and University,\\
  1088 Budapest, M\'uzeum krt.~6--8, Hungary}
\address{\ddag\ Department of Nonlinear and Quantum Optics,\\
  Institute for Solid State Physics and Optics,
  Hungarian Academy of Sciences,\\
  P.O.\ Box 49, H-1525 Budapest, Hungary}

\begin{abstract}
  A great number of problems of relativistic position in quantum
  mechanics are due to the use of coordinates which are not inherent
  objects of spacetime, cause unnecessary complications and can lead
  to misconceptions.  We apply a coordinate-free approach to rule out
  such problems.  Thus it will be clear, for example, that the Lorentz
  covariance of position, required usually on the analogy of Lorentz
  covariance of spacetime coordinates, is not well posed and we show
  that in a right setting the Newton--Wigner position is Poincar\'e
  covariant, in contradiction with the usual assertions.
\end{abstract}


\section{Introduction\label{sec:Intro}}

The position observable in relativistic quantum mechanics is an old
problem without a fully satisfactory solution; good summaries of the
question are given in Refs.~\cite{kalnay} and \cite{bacry}.  The
trouble is that there is no position observable which has all the
natural properties we expect on the base of nonrelativistic quantum
mechanics and, moreover, satisfies the requirement of covariance.
Earlier position was looked for as a family of Lorentz covariant
operators, then projection valued measures or positive operator valued
measures were investigated in a system of imprimitivity, and recent
publications deal with a collection of projections or positive
operators which are related to the structure of spacetime in a
convenient way \cite{fleming,bush}.

In usual treatments spacetime is always considered in coordinates.
``Much conceptualization in contemporary physics is bogged down by
unnecessary assumption concerning a specific choice of
coordinates\dots'' \cite{post} which results in needless complications
and can lead to conceptual errors, too.  For instance, it is false to
require Lorentz covariance of position observable on the analogy of
Lorentz covariance of spacetime coordinates (see Section
\ref{seq:Observables}).

In the present paper we put the problem of position observable into a
structure of \emph{Spacetime without reference frames} which
eliminates the irrelevant matters, throws new light on the old results
and admits new ones, too.

To illustrate the misleading feature of coordinates, let us recall
some usual statements regarding position observable.

1. ``The laws of physics should be invariant under transformations of
reference frames.  This symmetry is guaranteed by postulating the
existence of ten infinitesimal generators\dots'' of a unitary
representation of the Poincar\'e group \cite{jordan}.

It is the free particles that are classified by representations of the
Poincar\'e group: only closed systems have Poincar\'e symmetry.  The
equivalence of reference frames is independent of what is described, a
closed system or a not closed one.  If we use \emph{Spacetime without
reference frames}, then passive Poincar\'e transformations of
reference frames will be of no importance, while active Poincar\'e
transformations are the automorphisms of spacetime and become
symmetries of a free system. The confusion of active and passive
Poincar\'e transformations yields that one tries to impose the same
transformation rule on position coordinates as on the spacetime
coordinates.

2. ``\dots it would be difficult to conciliate the operator character
of position with the parameter character of time \cite{bacry}.''

The use of coordinates confuses some notions: there is spacetime,
there are (different) times and (different) spaces according to
(different) inertial observers; but position observable (with respect
to an observer), whatever it is, though being related to, is not equal
to the space of the observer in question.  We can define spacetime
position as a family of observables with respect to an arbitrary
observer $\uu$; these observables have a timelike component and a
spacelike component relative to an observer $\uu'$. The timelike
component is a c-number if and only if $\uu=\uu'$ (see Section
\ref{seq:Observables}).

3. The main objection to the Newton--Wigner \cite{newton-wigner}
position (besides that it is not Lorentz covariant) is that
``localization should also be Lorentz invariant'' but it turns out
that ``if a state is localized for one observer, it is no longer
localized for another one'' which contradicts Lorentz invariance
\cite{kalnay}.

Lorentz invariance does not mean that something must be the same for
all observers.  Let us consider a classical mass point: it can be at
rest with respect to an observer but this does not imply that it must
be at rest with respect to all observers.  Replacing ``at rest'' with
``localized'', we see that the statement ``if a state is localized for
one observer, it is no longer localized for another one'' does not
break Lorentz invariance (see Section \ref{seq:Causality}).

\section{Special relativistic spacetime model\label{sec:SpecRel}}

We shall use \emph{Spacetime without reference frames} introduced in
Ref.~\cite{whitebook} to investigate the problems of position
operator.  In such a framework, working with \emph{absolute objects},
i.e.\ with ones free of coordinates and distinguished observers, we
rule out questions regarding Lorentz covariance in the conventional
treatments.  Although the advantages of this model are well-known
\cite{model1,model2,model3,model4,model5}, a brief recapitulation of
its fundamental concepts is noteworthy.

In usual treatment, spacetime is considered to be $\RR\times\RR^3$.
While spacetime indeed \emph{can} be represented by $\RR\times\RR^3$,
it is also possible to work with less particular mathematical objects.
The physical meaning behind $\RR\times\RR^3$ is fixing an observer, an
origin and some coordinate axes.  Thus in the usual treatment what
really happens is the following: one defines the space \emph{and} the
time of an observer and then gives transformation rules to change
observers.  Spacetime as an affine space endowed with some further
structure (e.g.\ Lorentz form) can be well treated mathematically
without appealing to $\RR\times\RR^3$.  Instead of giving
transformation rules, we can define the notion of an observer and then
\emph{calculate} how things seem for different observers.

Let us now formalize the essence of this spacetime model and fix some
notations. Let $\MM$ be a four dimensional oriented real vector space,
while $M$ is an affine space over $\MM$, representing the set of
spacetime vectors and spacetime points, respectively.  Let $\II$ be a
one dimensional oriented real vector space: the measure line of
spacetime distances (thus for example the time unit sec is an element
of $\II$).  Although spacetime distances could be measured in real
numbers after fixing a unit, this would keep us away from talking
about the physical dimension of quantities in question.

Further let $\cdot: \MM\times\MM \to \II\otimes\II$ be symmetrical,
bilinear map of the type of 3 plus 1 minus (Lorentz product), endowed
with an \emph{arrow orientation} which determines the \emph{future
directed} timelike and lightlike vectors.  Note that the Lorentz
product of two spacetime vectors is an element of $\II\otimes\II$,
that is, it has the physical dimension of sec$^2$.

Many times division by time intervals occurs, e.g.\ in derivation of
velocity.  Such a procedure is handled properly through the use of the
tensorial quotients of vector spaces.  Thus an absolute velocity,
which is a spacetime vector over a time interval, is an element of
$\frac{\MM}{\II}$.  The Lorentz product can be naturally transferred
onto $\frac{\MM}{\II}$ where it will be real valued.

The set of absolute velocities is
\begin{equation*}
   V(1):=\left\{\uu\in\frac{\MM}{\II}
   \biggm| \mbox{$\uu\cdot\uu=-1$,
   $\uu$ is future directed}\right\}. 
\end{equation*} 
Given a $\uu\in V(1)$, we define
\begin{equation*}
   \Eu:=\{\,\xx\in\MM \mid
   \uu\cdot\xx=0\,\}
\end{equation*}
which is a three dimensional spacelike linear subspace of $\MM$.  The
restriction of the Lorentz product onto $\Eu$ is an $\II\otimes\II$
valued Euclidean product.

Every spacetime vector can be uniquely split into the sum of a
timelike vector parallel to $\uu$ and a spacelike vector in $\Eu$, in
other words, we can give the $\uu$-{\it splitting}
\begin{equation*}
   \MM\to\II\times\Eu, \quad
   \xx\mapsto (\btau_\uu(\xx),\bpi_{\uu}(\xx))
\end{equation*}
where
\begin{equation*}
   \btau_\uu(\xx):=-\uu\cdot\xx,\quad
   \bpi_{\uu}(\xx):=\xx-\btau_\uu(\xx)\uu. 
\end{equation*}

The best way to formalize our picture about an observer is to define
it to be a collection of world lines that satisfies some requirements
(e.g.\ no self-intersections). A point of the space of an observer is
in fact a world line.  An inertial observer is an observer with only
straight, parallel world lines; thus an inertial observer can be given
by an absolute velocity $\uu\in V(1)$.  According to Einstein's
synchronization, spacetime points $x$ and $y$ are
\emph{$\uu$-simultaneous} if and only if $\uu\cdot(x-y)=0$, in other
words, $x-y \in \Eu$.  Thus $\uu$-simultaneous spacetime points form
an affine hyperplane over $\Eu$.  A $\uu$-simultaneous hyperplane is
considered to be a \emph{$\uu$-instant} and the set $I_\uu$ of such
hyperplanes is the time of the observer, briefly the
\emph{$\uu$-time}.  The time interval between $\uu$-instants $t_1$ and
$t_2$ is defined to be
\begin{equation*}
   t_1-t_2:=\btau_\uu(x_1-x_2)\qquad
   (x_1\in t_1,x_2\in t_2)
\end{equation*}
which is a good definition as it is independent of the choice of $x_1$
and $x_2$.  $I_\uu$ endowed with this subtraction is an affine space
over $\II$.

The space points of the inertial observer $\uu$ are straight lines in
spacetime, parallel to $\uu$.  The space of the observer $\uu$,
denoted by $E_\uu$, endowed with the subtraction
\begin{equation*}
   q_1-q_2:=\bpi_\uu(x_1-x_2)
   \qquad (x_1\in q_1, x_2\in q_2)
\end{equation*} 
is an affine space over the vector space $\Eu$ (the definition is
independent of the choice of $x_1$ and $x_2$).

The Lorentz group is
\begin{equation*}
   \mathcal L:=\{\, \LL:\MM\to\MM \mid
   \mbox{$\LL$ is linear, $\LL\xx \cdot \LL\mathbf y
   =\xx\cdot\mathbf y$ ($\xx, \mathbf y \in\MM$)} \,\}. 
\end{equation*}

Orthochronous Lorentz transformations preserve the arrow orientation
of the Lorentz form.

\emph{The} three-dimensional orthogonal group is not a subgroup of the
Lorentz group (contrarily to the usual statement in the coordinatized
treatment).  For all $\uu\in V(1)$,
\begin{equation*}
   {\mathcal O}_\uu:= \{\, \LL\in{\mathcal L} 
   \mid \LL\uu=\uu \,\}
\end{equation*}
is a subgroup of the Lorentz group which is isomorphic to the
three-dimensional orthogonal group (in fact the restrictions of the
elements of $\mathcal O_\uu$ onto the three dimensional Euclidean
space $\Eu$ are orthogonal maps).  $\mathcal O_\uu$ and $\mathcal
O_{\uu'}$ are different if $\uu\neq\uu'$.

Similarly, \emph{the} time inversion and \emph{the} space inversion
are not elements of the Lorentz group.  For all $\uu\in V(1)$ we can
give the $\uu$-time inversion and the $\uu$-space inversion:
\begin{equation*}
   \xx \mapsto-\btau_\uu(\xx)\uu + \pi_\uu(\xx),\quad
   \xx \mapsto \btau_\uu(\xx)\uu - \pi_\uu(\xx).
\end{equation*}

The Poincar\'e group is
\begin{equation*}
   {\mathcal P}:=\{\, L:M\to M \mid
   \mbox{$L$ is affine, $\LL\in\mathcal L$}\,\}
\end{equation*}
where $\LL$ denotes the linear map under $L$.  A Poincar\'e
transformation over an orthochronous Lorentz transformation is called
orthochronous.

The Lorentz group is not a subgroup of the Poincar\'e group
(contrarily to the usual statement in the coordinatized treatment); it
cannot be, since Lorentz transformations are $\MM\to\MM$ linear maps,
Poincar\'e transformations are $M\to M$ affine maps.  For all $o\in
M$,
\begin{equation*}
   {\mathcal L}_o:=\{\, L\in {\mathcal P}
   \mid L(o)=o \,\}
\end{equation*}
is a subgroup isomorphic to the Lorentz group, but ${\mathcal L}_o$
and ${\mathcal L}_{o'}$ are different for different $o$ and $o'$. The
elements of $\mathcal L_o$ are called $o$-homogeneous Poincar\'e
transformations.

Of course, neither \emph{the} time inversion nor \emph{the} space
inversion are elements of the Poincar\'e group. We can only define a
time inversion with respect to an observer $\uu$ and a time a
$\uu$-instant $t$.

For all $\uu\in V(1)$ and $t\in I_\uu$
\begin{equation*}
   {\mathcal E}_{\uu,t}:=\{\, L\in \mathcal P 
   \mid L[t]=t \,\}
\end{equation*}
is a subgroup of the Poincar\'e group; the restriction of its elements
onto $t$ are Euclidean transformations of the hyperplane $t$;
moreover, it contains the $\uu$-time inversion with respect to the
$\uu$-instant $t$.

\section{Position observable(s)\label{seq:Observables}}

Most of today's quantum physics starts with giving the following
objects associated with the physical system: a Hilbert space and a
(unitary ray) representation of the automorphism group (symmetries) of
the used spacetime model on it.  Pure states of the system then
realized as rays of the Hilbert space.  There are different possible
interpretations of these mathematical objects.  It is common for
example to think of a state as something changing by time, i.e.\ a
time dependent ray.  However, in absolute description we can not talk
about ``time evolution'' (who's time?) and so we have to use another
picture.  In absolute description a system does not go through an
evolution by time, it simply \emph{exists} in spacetime. An observable
at a certain time instant, however, is conceptually different in the
absolute description of the ``same'' observable at a different time
instant and thus we represent them by two not necessarily identical
operators.

One should also take note of the fact that in absolute description
passive spacetime transformations (change of coordinate system) are of
no importance; we emphasize that the representation of the Poincar\'e
group corresponding to a closed system does not refer to the
equivalence of reference frames, thus it has nothing to do with that
``the laws of physics should be invariant under transformations of
reference frames".  The representation reflects the properties of the
physical system in question, namely that the particle is free; we
think of a spacetime symmetry as a transformation that---in case of a
closed system---turns a possible \emph{process} (``a full time
evolution of the system'') into another possible process of the
system, i.e.\ that maps the set of pure states into itself.

A convenient way to describe physical quantities like position is to
use projection valued measures or positive operator valued measures.
Wightman \cite{wightman} defined localization, i.e.\ position of a
free particle as a projection valued measure $P$ defined on the Borel
subsets of space such that $U_S P(E) U_S^{-1}=P(S[E])$ for all Borel
subsets $E$ of space and for $S$ being an arbitrary Euclidean
transformation in space or the time inversion, where $U$ is the
corresponding representation of the Poincar\'e group.

Since neither \emph{the} space nor \emph{the} Euclidean subgroup of
the Poincar\'e group nor \emph{the} time inversion exist, we
reformulate this approach in our framework as follows.

Consider an observer $\uu$ and a $\uu$-instant $t$. For every Borel
set $E \in {\mathcal B}(t)$ there should be a projection
$P_{\uu,t}(E)$ standing for the event of the particle being located in
$E$.  By the natural expectations of localization, $P_{\uu,t}$ is
required to be a projection valued measure having the following
connection with the representation of the Poincar\'e group.
\begin{equation}\label{eq:CovTran}
   U_S P_{\uu,t} (E) U_S^{-1} 
   = P_{\uu,t} (S[E])
\end{equation}
for all $E \in {\mathcal B}(t)$ and $S \in {\mathcal E}_{\uu,t}$.
Since we only want to deal with a one particle system, in the
following we will always consider an irreducible representation of the
Poincar\'e group.

Applying Wightman's proof, we can state that for fixed $\uu$ and $t$,
a projection valued measure satisfying (\ref{eq:CovTran}) is unique
under some regularity conditions.

Note that we have \emph{many} spacelike hypersurfaces, and of course,
localization on one of them is not the same as on another one.
Furthermore, the transformation rule (\ref{eq:CovTran}) says nothing
about the relation between $P_{\uu,t}$ and $P_{\uu',t'}$ for
${\uu'}\neq\uu$ or $t'\neq t$.  Nevertheless, the following nice
transformation property can be shown:

\begin{prop}
  Let an imprimitivity system (\ref{eq:CovTran}) be given for all
  $\uu\in V(1)$ and $t\in I_\uu$. If Wightman's regularity condition
  holds then
  \begin{equation}\label{eq:covtran}
     U_L P_{\uu,t}(E) U_L^{-1}
     = P_{\LL\uu,L[t]}(L[E])
  \end{equation}
  for all $\uu\in V(1)$, $t\in I_\uu$, Borel subset $E$ of $t$ and for
  all orthochronous Poincar\'e transformations $L$.
\end{prop}

\begin{proof}
  Let $L$ be fixed; putting
  \begin{equation*} 
     \overline P_{\uu,t}(E) := U_L^{-1}
     P_{\LL\uu,L[t]}(L[E]) U_L
  \end{equation*}
  we find that $\overline P_{\uu,t}$ satisfies (\ref {eq:CovTran}),
  since if $S\in{\mathcal E}_{\uu,t}$ then $LSL^{-1}$ is in ${\mathcal
  E}_{\LL\uu, L[t]}$.  As a consequence of uniqueness, we have the
  desired result.
\end{proof}

It is known that integrating the space coordinates by Wightman's
projection valued measure, one gets the Newton--Wigner position.

Accordingly, by choosing a spacetime origin $o$, with the aid of the
above projection valued measure we can construct a family of position
operators:
\begin{equation*}
   W_{\uu,t}^o := \int_t
   (\mathrm{id}_t -o) \, dP_{u,t} \quad
   (o\in M, {\uu} \in V(1), t \in I_{\uu}). 
\end{equation*}
$W_{\uu,t}^o$ is an $\MM$ valued totally self-adjoint vector operator
which we call the $o$-centered generalized Newton--Wigner position at
the ${\uu}$-instant $t$.

Using the transformation properties of integration by projection
valued measure we can easily find the transformation rule of the
members of the family of generalized Newton--Wigner positions:

\begin{prop}~
  \begin{equation}\label{eq:NWtraf}
     U_L {W_{\uu,t}^o} U_L^{-1}
     = \mathbf L^{-1}W_{\mathbf L\uu,L[t]}^{Lo}
  \end{equation}
\end{prop}

We now understand that the above equality is the Poincar\'e covariance
of the generalized Newton--Wigner position.  We emphasize that this
Poincar\'e covariance of the family of positions does not refer to the
equivalence of reference frames; it reflects the properties of the
particle according to what has been said in the beginning of the
current Section.

It is important to see that $W_{\uu,t}^o$ is a ``four-vector''
($\MM$-valued) but it does not transform as a spacetime-vector, i.e.\
for a fixed $\uu$, $t$ and spacetime origin $o\in t$ (which
corresponds to the usual considerations in coordinates), $Q :=
W_{\uu,t}^o$ is not a ``four-vector operator'': $U_L^{-1} Q U_L^{1}
{\neq} \LL Q$ for an $o$-homogeneous Poincar\'e transformations $L$.

The $\uu$-spacelike component of $W^o_{\uu,t}$ corresponds to the
original Newton--Wigner position. It is interesting, however, that we
can consider its $\uu'$-spacelike components, too.  Applying
(\ref{eq:NWtraf}), we easily find:

\begin{prop}
  The $\uu'$-spacelike component of $W_{\uu,t}^o$ transforms as a
  $\uu'$-spacevector, that is,
  \begin{equation*}
     U_L \bpi_{\uu'}(W_{\uu,t}^o) U_L^{-1} 
     = \mathbf R^{-1} \bpi_{\uu'}(W_{\uu,t}^o)
  \end{equation*}
  for $o$-homogeneous $L \in \mathcal E_{\uu',t'}$ if and only if
  $\uu=\uu'$, where $\mathbf R$ is the restriction of $\mathbf L$ onto
  $E_{\uu'}$ (a rotation in $\mathbf E_{\uu'}$).
\end{prop}

The generalized Newton--Wigner position has timelike component, too,
for which we derive the following interesting result.

\begin{prop} 
  The $\uu'$-timelike component of $W_{\uu,t}^o$ is a c-number if and
  only if $\uu=\uu'$.
\end{prop}

\begin{proof}
  Using the properties of integration of projection valued measures,
  it is easy to see that the $\uu'$-timelike component is a c-number
  if and only if $\btau_\uu (\mathrm{id}_t-o)$ is constant almost
  everywhere according to $P_{\uu,t}$.  It is constant only on the
  two-dimensional affine subspaces of $t$ parallel to $\Eu\cap\mathbf
  E_{\uu'}$.  But considering the transformation rules
  (\ref{eq:covtran}), it is impossible that the support of $P_{\uu,t}$
  is in one of these subspaces.
\end{proof}

\section{Localization and causality\label{seq:Causality}}

Let us investigate localization problem in our framework.  We conceive
that a state $\Phi$ (i.e.\ an element of the Hilbert space) is
\emph{localized} in a set $E\in\mathcal B(t)$ at a $\uu$-instant $t$
if $P_{\uu,t} (E) \Phi = \Phi$ holds.

Poincar\'e invariance of localization means that if $\Phi$ is
localized in $E$ at a $\uu$-instant $t$ and $L$ is a proper Poincar\'e
transformation then $U_L\Phi$ is localized at the $\LL\uu$-instant
$L[t]$ in $L[E]$, which trivially holds.

Now it is clear that the requirement of Lorentz invariance ``if a
state is localized for one observer, it must be localized for all
other ones'' is not well posed, Lorentz invariance---or better,
Poincar\'e invariance---should mean that {\it if a state is localized
for one observer then a Poincar\'e transform of the state must be
localized for the corresponding transformed observer}.

By causality, we expect that if $\Phi$ is localized in $E \in
{\mathcal B}(t)$ then $\Phi$ is localized in $(E+T) \cap t' \in
{\mathcal B}(t')$, that is, $P_{\uu',t'}((E+T) \cap t') \Phi=\Phi$
holds for every observer $\uu'$ and $\uu'$-instant $t'$, where $T$
denotes the cone of timelike vectors.

{\it The existence of a state localized for one observer and not
localized for another one}, i.e.\ the existence of a $\phi$ such that
$P_{\uu,t}(E)\Phi=\Phi$ for a $t\in I_\uu$ but $P_{\uu',t'}((E+T)\cap
t')\Phi\neq\Phi$ for some $\uu'$-instant $t'$ {\it denies causality
but not the Poincar\'e invariance}.

The acausal feature of the Dirac equation is well known and thoroughly
treated in the literature \cite{heger-ruijsen,ruijsen,heger}.

The causality requirement yields that $P_{\uu,t}(E)$ and $P_{\uu',
t'}(E')$ are orthogonal if $E$ and $E'$ are spacelike separated. It is
known that projection valued measure satisfying covariant
transformation rules and local commutativity ($[P_{\uu,t}(E),
P_{\uu',t'}(E')] = 0$) is equal to zero \cite{bush,malament}. That is
why the generalized Newton--Wigner position violates causality, though
being Poincar\'e covariant.

\section{Discussion\label{sec:discus}}

In the present paper we have investigated an old problem in
relativistic quantum mechanics: to find position operator which has
natural properties expressed in transformation rules. On the other
hand projection or positive operator valued measure facilitated to
express our expectations on the notion of localization according to
our intuitive picture. A fundamental result was Wightman's statement
about uniqueness of a projection valued measure describing
localization. It seemed to be worth paying attention not only to its
projection decomposition but the operator itself, too.

In the current paper we have used a special relativistic spacetime
model free of distinguished observers and reference frames. With the
aid of this formalism it is obvious how physical quantities like
position are connected to observers of spacetime.

For different observers, position corresponds to localization on
different, not even parallel hypersurfaces; and for a single observer
but different time instants it corresponds to localization on parallel
but still not equal hypersurfaces (this is because position is not a
constant of motion).  Therefore, instead of a \emph{single} position,
we have a family of position operators, the generalized Newton--Wigner
position, labelled by observers and time instants (and spacetime
origins), which is Poincar\'e covariant.  Each member of the family is
an $\MM$ valued vector operator whose spacelike and timelike
components behave different for different observers.

\end{document}